\documentclass{aa}
\authorrunning{P. Saracco et al.}
\input psfig.tex
\begin{document}

   \thesaurus{11     
              (12.03.3;  
               11.05.2;  
               11.12.2;
	       11.16.1;  
               11.19.7;
		13.09.1)} 
   \title{IR Colors and Sizes of Faint Galaxies\thanks{Based on
   observations collected at the European Southern Observatory, La
   Silla, Chile}}


   \author{P. Saracco\inst{1},
          S. D'Odorico\inst{2},
	A. Moorwood\inst{2},
	A. Buzzoni\inst{1},
	J.- G. Cuby\inst{3},
	C. Lidman\inst{3} 
}

   \offprints{Paolo Saracco}

   \institute{$^1$ Osservatorio Astronomico di Brera, via E. Bianchi 46,
22055 Merate (LC), Italy \\
	$^2$ European Southern Observatory, ESO, Karl-Schwarzshild Strass 2,
D-85748, Garching bei M\"{u}nchen, Germany\\
$^3$  European Southern Observatory, ESO, Casilla 19001, Santiago 19, Chile
             }

   \date{Received ......, 1999; accepted }

   \maketitle

   \begin{abstract} We present J and Ks band galaxy counts down to
J=24 and Ks=22.5 obtained with the new infrared imager/spectrometer,
SOFI, at the ESO New Technology Telescope.  The co-addition of short,
dithered, images led to a total exposure time of 256 and 624 minutes
respectively, over an area of $\sim20$ arcmin$^2$ centered on the NTT
Deep Field.  The total number of sources with S/N$>5$ is 1569 in the J
sample and 1025 in the Ks-selected sample.  These are the largest
samples currently available at these depths.  A d$logN$/d$m$ relation
with slope of $\sim0.36$ in J and $\sim0.38$ in Ks is found with no
evident sign of a decline at the magnitude limit.  The observed
surface density of ``small'' sources is much lower than ``large'' ones
at bright magnitudes and rises more steeply than the large sources to
fainter magnitudes. Fainter than $J\sim22.5$ and Ks$\sim21.5$, small
sources dominate the number counts. Galaxies get redder in J-K down to
 J$\sim20$ and Ks$\sim19$.  At fainter magnitudes, the median
color becomes bluer with an accompanying increase in the compactness
of the galaxies.  We show that the blue, small sources which
dominate the faint IR counts are not compatible with a high redshift
($z>1$) population.  On the contrary, the observed color and
compactness trends, together with the absence of a turnover at faint
magnitudes and the dominance of small sources, can be naturally
explained by an increasing contribution of sub-$L^*$ galaxies when
going to fainter apparent magnitudes.  Such evidence strongly supports
the existence of a steeply rising ($\alpha\ll-1$) faint end of the
local infrared luminosity function of galaxies - at least for
luminosities $L<0.01L^*$.

               
   \end{abstract}

%

\section{Introduction}
Near infrared (IR) selected samples may provide significant advantages
over optically selected samples in studying galaxy evolution due to
the small and almost galaxy type independent $k$-corrections at these
wavelengths (Cowie et al. 1994).  Moreover, IR selection provides
samples which are not biased towards star-forming galaxies and may
trace the mass of the galaxies over a wide range of redshift.  For
these reasons, since the beginning of nineties, near IR galaxy counts
have been considered a powerful cosmological test.  This, however, was
discovered not to be the case when it was realized that deep optical
and near IR observations were difficult to reconcile without  assuming
an {\em ad hoc} population of galaxies or extreme merging scenario (Cowie
et al. 1990; Babul \& Rees 1992; Broadhurst et al.  1992; Gronwall \&
Koo 1995; Yoshii \& Peterson 1995).  Moreover, the large uncertainty
on the slope of the faint-end tail of the local LF, especially in the
near IR, means that the hypothesis on the nature of the very faint
population of galaxies is completely unconstrained in terms of redshift
and thus in terms of evolution.

The addition of color information to number counts may shed light on the
different populations that  contribute at a given magnitude and/or
the  redshift of galaxies beyond the spectroscopic
limits (e.g. Lilly, Cowie and Gardner 1991).

Some efforts have been made to include in the analysis of faint
galaxies also information about the morphological structure
derived both from ground based observations (e.g. Lilly et al. 1991; Colless
et al. 1994) and, more recently, from HST images.

Morphological classifications based on visual inspection of  optical HST images
have  revealed that the faint field population is apparently dominated
by irregular and merging galaxies (Glazebrook et al. 1995; Driver et al. 1995).
However, the comparison of optical images of distant galaxies with possible
local counterparts could give misleading results since the surface
brightness dimming and the reduced signal-to-noise ratio tend
to emphasize high contrast features.
Also, as discussed by  Abraham
et al. (1996), the band-pass shifting effects arising from the k-correction
implies that low-redshift optical images are compared with high-redshift 
UV images which, inevitably, show an increasing irregularity in their
morphology.

Most of the uncertainties related to  image structure  vanish when
dealing with near-IR data since they trace the underlying old
stellar population of galaxies  out  to $z\sim3$-4.

Recently, Bershady et al. (1998) derived deep near-IR galaxy counts 
 in two high-latitude Galactic fields, summing up to 1.5 arcmin$^2$, 
observed with the Keck telescope to depths of J=24.5 and K=24.
They find that counts do not roll over by K$\sim22.5$ and that the  smallest 
galaxies dominate at magnitudes fainter than J$\sim23$ and K$\sim21.5$.
They also find  no hint of a flattening in the counts at the
limits of the survey and mean colors bluer than  expected 
from a no evolution prediction.
These  features are discussed under the hypothesis of 
an increasing contribution of sub-L$^*$ galaxies at faint magnitudes,
i.e. a rising faint end luminosity function (LF), and/or evolution
of galaxies.

In this paper we present counts, near-IR colors, sizes and compactness indices
of faint galaxies observed at the ESO NTT telescope to depths
comparable with the faintest previous ground-based surveys but over a much
larger area. In \S 2 we describe the
observations and the photometric calibration while in \S 3 we describe
the image processing and analysis and the construction of our
photometric sample.  In \S 4 the differential galaxy counts and colors
are derived while in \S 5 we describe the method used to estimate the size
and compactness of sources and present counts as a function of
galaxy size.  The discussion of the results is presented in \S 6 and
section \S 7 summarizes our results and conclusions.

\section{Observations and Photometric Calibration }
The observational data used were obtained during commissioning of the
SOFI infrared imager/spectrometer (Moorwood et al. 1998) at the ESO
3.5 m New Technology Telescope (NTT) in March 1998 and are publicly available
(http://www.eso.org/science/sofi\_deep/). SOFI is
equipped with a 1024$\times$1024 pixel Rockwell Hawaii array providing
a plate scale of 0.292$\pm 0.001$ arcsec/pix in the wide-field
observing mode.  The total field of view on the sky is thus about
5$\times$5 arcmin per frame.  Field center was located at
$\alpha=12^h:05^m:26.02^s$, $\delta=-07^o:43':26.43''$ and
observations were made in the two near-infrared filters J
($\lambda_c$=1.247 $\mu$m; $\Delta\lambda\sim0.290 $ $\mu$m ), and Ks
($\lambda_c$=2.162 $\mu$m;$ \Delta\lambda\sim0.297$ $\mu$m).

The observations were gathered over several nights but under quite
homogeneous seeing conditions ranging from 0.6-0.9 arcsec (median FWHM
= 0.75 arcsec). The standard jitter technique was used with effective
exposure times for each individual frame of 1 min in Ks, (resulting
from the average of 6 exposure of 10 s each), and 2 min in J
(resulting from the average of 6 exposures of 20 s each).The jitter
offsets were controlled by an automated template procedure (the reader
is referred to the SOFI User Manual for more details) which was used
to randomly offset the telescope position within a box of 40 arcsec
on the sky.  The deepest exposures were therefore achieved across the
central 4.45$\times$4.45 arcmin field. A single measurement normally
consisted of 60 and 30 frames in Ks and J, respectively, the whole
``duty'' cycle amounting to a total of 1 hour integration. These were
repeated to achieve total exposure times of 624 min in Ks and 256 min
in J in the final co-added images.
\begin{table*}
\caption{ Summary of observations and image quality.}
\begin{center}

 \begin{tabular}{cccccc}

\hline
\hline
        Filter &  Number of & t$_{exp}$& $<FWHM>$& $\mu$&m$_{lim}$\\
        & frames    &  (s)   &(arcsec)&mag/arcsec$^2$& (3$\sigma$)\\
\hline
          Ks &  624 & 37440 & 0.75& 23.95&22.76\\
          J  &  128 & 15240 & 0.75& 25.75&24.57\\
\hline
\hline
\end{tabular} 
\end{center}
\end{table*}

Photometric calibration of the observations has been made 
by observing 
several standard stars from  the list of Infrared NICMOS Standard Stars 
 (Persson et al. 1998).  
Star magnitudes ranged between $10.8<Ks<11.7$ and 
$11.3<J<12.0$, and instrumental total flux has been estimated 
by deriving the growth curve for each star. 
This allowed us to achieve a typical uncertainty of $\pm 0.02$ mag
in the photometric zero point.

The estimated magnitudes have been then corrected for atmospheric extinction 
assuming A$_{Ks}$=0.11 and A$_J$=0.08. 

\section {Image Processing and Analysis}
\subsection{Flat Fielding}
Raw frames have been  first corrected for the bias  pattern by subtracting,  
for each night, a median dark frame and then flat fielded to 
correct for pixel-to-pixel gain variations.

Two different types of flat field were compared for each band.  A
differential dome flat field (DDF), obtained by subtracting lamp-on
and lamp-off images of the illuminated dome, and a ``superflat'' (SF)
obtained for each night by combining all the images of the observed
field (over 120 frames throughout).  Template flat-fields were
cleaned of both ``hot'' and ``cold'' pixels via the IRAF CCDMASK
routine (we adopted a $\pm6\sigma$ rejection threshold).  Rejected
pixels were ``filled in'' by local interpolation on the frame (via the
IRAF/FIXPIX routine), and the resulting final flats have been then
normalized to their mean value.

Comparison among the different flat fields has been made  on
the basis of both high- and low-frequency systematics, that is
including pixel-to-pixel variations as well as the magnitude scatter
of  star measurements made on a grid of positions across the
field. Having verified that both the DDF and SF flat fields gave
comparable results, allowing a magnitude accuracy within $\sim0.02$
mag, we decided to adopt the DDF as a fiducial template
for our observations.

\subsection{Co-addition}
After flat fielding and standard sky subtraction, each basic set of  frames 
was  registered and finally co-added to produce a single image 
corresponding to  one hour exposure. 
The whole procedure was accomplished using  the
original {\it Jitter} programme by Devillard 
(1998; http://www.eso.org/eclipse).

For each input frame, the software first generates a sky-subtracted
image resulting from filtering a set of typically 10 time-adjacent
frames.  Frame registering is then performed via cross-correlation
among the brightest objects in the fields. A sub-pixel offset accuracy
is assured by {\it Jitter} so that a re-sampling of the images is also
required before the stacking procedure.  Co-addition is then
performed, adding a filter to remove spurious pixel values before
averaging.

Since observations span different nights and were therefore collected
under different photometric conditions and airmass, each one hour
exposure image, derived from the recombination of the basic set of
frames, has been shifted to the same magnitude zero point before
co-addition to produce the final image.  In order to estimate the flux
scaling factor, the instrumental magnitudes of the 20 brightest
objects ($14<Ks<$19; $16<J<20$) were compared throughout the
frames.  The photometric uncertainty introduced by such scaling is
about $0.02$ mag.  Once rescaled, the 11 Ks frames and the 5 J frames
were co-added by {\it Jitter} without applying any
filtering before averaging.

The surface brightness cutoff as well as the 3$\sigma$ magnitude limit
for a 2$\times$ FWHM (1.5 arcsec) circular aperture from the co-added
Ks and J band frames are reported in Tab. 1.

\subsection{Object Detection}
Systematic object detection has been performed across the central 
deepest portion 
(4.45$\times$4.45 arcmin) of the images using the SExtractor analysis package 
 (Bertin \& Arnouts 1996).

Before running the search algorithm we first smoothed the final frames
by convolving with a 2.5 px FWHM ($\sim0.73$ arcsec) Gaussian PSF,
picking up those objects exceeding a 1$\sigma_{bkg}$ threshold over
the background RMS.  This threshold corresponds to
2.2$\sigma_{bkg}/\sqrt{N}$, N being the minimum detection area in
pixels i.e the number of connected pixels within one seeing disk.

This rather low detection threshold provided a raw catalog from which
we extracted different samples by varying the S/N limit in order to
optimize completeness and minimize the number of spurious detections
for a given magnitude limit.  For example, at a $S/N=5$ detection
limit, compact objects as faint as $Ks = 23.0$ and $J = 24.8$ were
detected in our images, compared with $Ks = 23.9-24.5$ and $J =
24.9-25.2$ as in the deepest Keck observations (Bershady et al. 1998;
Djorgovski et al. 1995). We are therefore confident that our sample is
not biased against the detection of small compact sources and that it
is fully comparable with the deepest data previously obtained by other
authors over smaller areas on the sky (Gardner et al. 1993; Soifer et
al. 1994; McLeod et al. 1995; Djorgovski et al. 1995; Moustakas et
al. 1997; Minezaki et al. 1998; McCracken et al. 1999). On the other
hand, the isophotal photometry at the adopted surface brightness limit
as reported in Tab. 1 directly affects the detection of extended
sources and low surface brightness objects and is fully comparable with the
data of Bershady et al. (1998) in their Keck images (i.e. K=23.9-24.2
mag/arcsec$^2$) or even deeper  (J=25-25.5 mag/arcsec$^2$).

\subsection{Magnitudes} 
\begin{figure}
\centerline{\psfig{figure=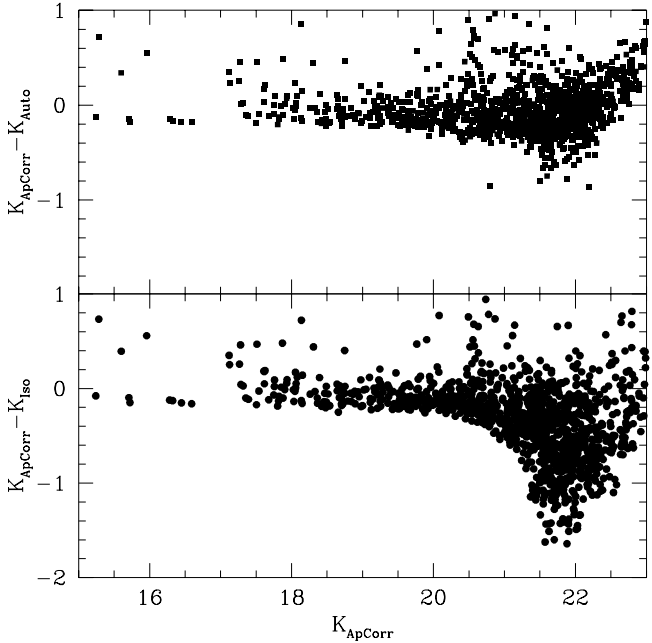,height=90mm}}
 \caption{Differences between the magnitude estimates for the Ks selected
sample:(upper panel) difference between aperture corrected magnitude 
($K_{ApCorr}$) and   AUTO (Kron-like) magnitude ($K_{Auto}$); (lower panel) 
difference between aperture corrected magnitude and isophotal magnitude 
($K{_Iso}$). 
The aperture correction (0.25 mag) between $2.5''$ and $5''$ 
has been obtained using a sample of sources brighter than Ks$=18.5$.}
\end{figure}
Aperture photometry over a 2.5 arcsec circular diaphragm
($\sim3\times$FWHM) has been derived for the whole sample and then
offset by 0.25 mag both in J and Ks in order to account for their
``total'' flux according to the mean growth curve as for the brightest
objects (Ks$<18.5$, J$<20$).  In Fig.1 the
corrected aperture photometry is compared with the
AUTO (Kron-like) magnitudes (Kron 1980) and isophotal magnitudes as
calculated by SExtractor for the Ks-band selected sample.

A significant deviation is found between the three sets of magnitudes.
Isophotal photometry tends to systematically underestimate the flux of 
faint objects,
an effect which is significant also at bright magnitudes (Ks$\sim20.5$).
To a lesser extent, an underestimate is also present in the case of the 
AUTO magnitude, especially in the two faintest magnitude
bins (Ks$>21.5$) where the background fluctuations become comparable
to the signal even at a few pixels from the center of the sources.

\subsection{Sample Selection}
On the basis of the adopted detection parameters (cf. \S 4.1), 1271 and 1743 
sources have been recognized by SExtractor in the Ks and J images,
respectively.

A measure of the detection reliability is now necessary in order to 
evaluate the number of spurious sources included in the sample, and 
their magnitude distribution. 
Since the completeness level in the detection algorithm 
is a function of object magnitude and size, our test will provide a 
guess of the optimum S/N ratio to eventually extract the fiducial J 
and Ks samples.

Operationally, we created a homogeneous set of noise images both in the J
and Ks bands by combining unregistered frames.  This assures a
similar background as in the original data.  The noise frames have
then been ``reversed'' (i.e. multiplied by -1) in order to reveal
the negative fluctuations and to make negative (i.e. undetectable)
possible residuals of real objects.  By running SExtractor with the
same detection parameter set used to search for sources in the science
frames, we have obtained the number of false detections which affect
our data.  The J band image is characterized by a very low noise and
most of the spurious detections occur at a S/N$<$2.5 level.  The Ks
band image displays on the contrary a larger noise which consequently
increases both the number of false detections as well as their S/N.
In Fig.2 we show the false detection distribution as a function of the
Ks magnitude.  From the figure, one sees that spurious detections
would significantly  affect photometry at low S/N, and a cutoff about
S/N$\sim$5 seems to ensure the optimum threshold for confident object
detection both in J and Ks within a 0.3 mag accuracy.  In particular,
using this cutoff the J selected sample is completely free from
spurious detections and the K-selected one down to K=21.5, while they
 become $\sim10\%$ at 21.5$<Ks\le22.5$.
\begin{figure}
\centerline{\psfig{figure=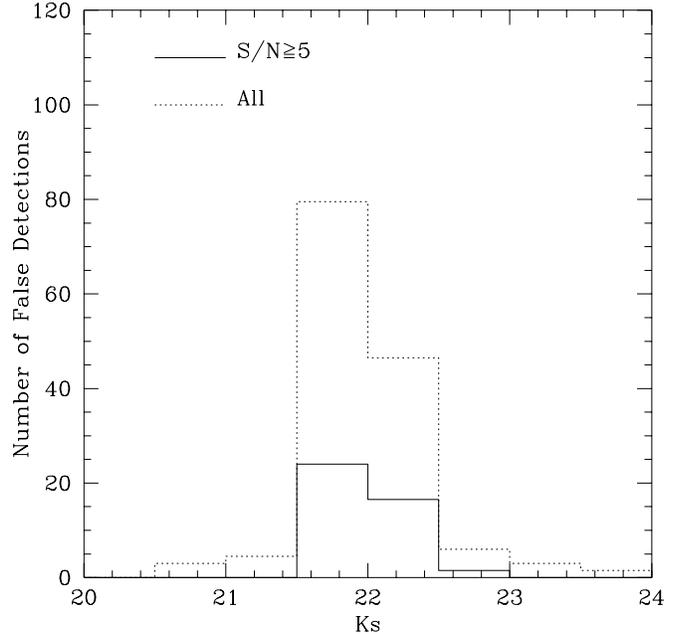,height=90mm}}
 \caption{Distribution of spurious sources. The dotted histogram
refers to the distribution of detections with minimum S/N$=2.2$
within the seeing disk. The thick line describe the distribution 
of detections having S/N$\ge5$.}
\end{figure} 
On the basis of the above selection criterium our subsequent analysis
will rely on the {\it bona fide} Ks and J samples containing 1025
sources down to Ks$\le$22.5 and 1569 sources down to J$\le$24.0.
These samples are used throughout the rest of this paper.

\section{Number Counts and Colors}
\subsection{Differential Galaxy Counts}
In order to derive  differential galaxy counts, we have first ``cleaned'' 
the two samples of stars. 
This has been done relying on the SExtractor morphological
classifier. Stars were  defined as those sources  either brighter 
than Ks$=19$ or  J$=20$, having a value of the ``stellarity'' index larger 
than 0.9 in both photometric bands.

Within the quoted magnitude limits, we verified that the J and Ks
SExtractor classification coincides at a 95$\%$ confidence level while
some discrepancy arises at fainter magnitudes where, however, the star
contamination is negligible.  It is worth noting that the adopted
procedure tends to slightly overestimate the resulting galaxy counts
both at faint magnitudes, where no count correction was applied, and
at bright magnitudes where some fuzzy stars in the field could escape
the cleaning procedure.  On the other
hand, our choice ensures that the galaxy samples are not biased
against compact galaxies and that stellar samples are not contaminated
by unresolved galaxies.

The second step in deriving galaxy counts concerns the completeness correction
to our sample for faint, undetected galaxies. Obviously, this correction is 
larger at faint magnitudes and, in general, mainly depends on the source
spatial structure.
Such dependence is usually taken into account in the literature by introducing 
simulated sources with different typical profiles and magnitudes embedded in a 
Gaussian noise (e.g. Minezaki et al. 1998; Metcalfe et al. 1995) 
or in real frames (e.g. Saracco et al. 1997).
These two methods suffer from the  approximations introduced by the use
of the ``typical'' profiles which cannot fairly reproduce the manifold
of galaxy shapes.

A better approach would be to add to a real frame a grid of test
sources artificially dimmed (e.g. Arnouts et al. 1999; Bershady et al.
1998; Yan et al. 1998).  Template sources could be singled out from
the brightest sources in the sample but a problem is that they might
not conveniently span the whole range of size and shape of the faint
galaxies.  As a possible way out, we chose to generate a set of frames
by directly dimming the final J and Ks frames themselves by various
factors while keeping constant the background noise.  The advantage of
this procedure is of course to provide a fair artificial sample in the
real background noise.  SExtractor was then run with the same
detection parameters to search for sources in each dimmed frame.

The correction factor $c$ is the number of dimmed galaxies which
should enter the fainter magnitude bin over the number of detected ones.
Since we have selected the two samples at a S/N$\ge5$ level,
the correction factors refers to the number of such detections.
By relaxing the selection criterion, using for example  S/N=3,
the factor $c$ would decrease and it would be $\sim2$ (i.e. a
50$\%$ completeness) in the faintest magnitude bin. 

\begin{figure*}
\centerline{\psfig{figure=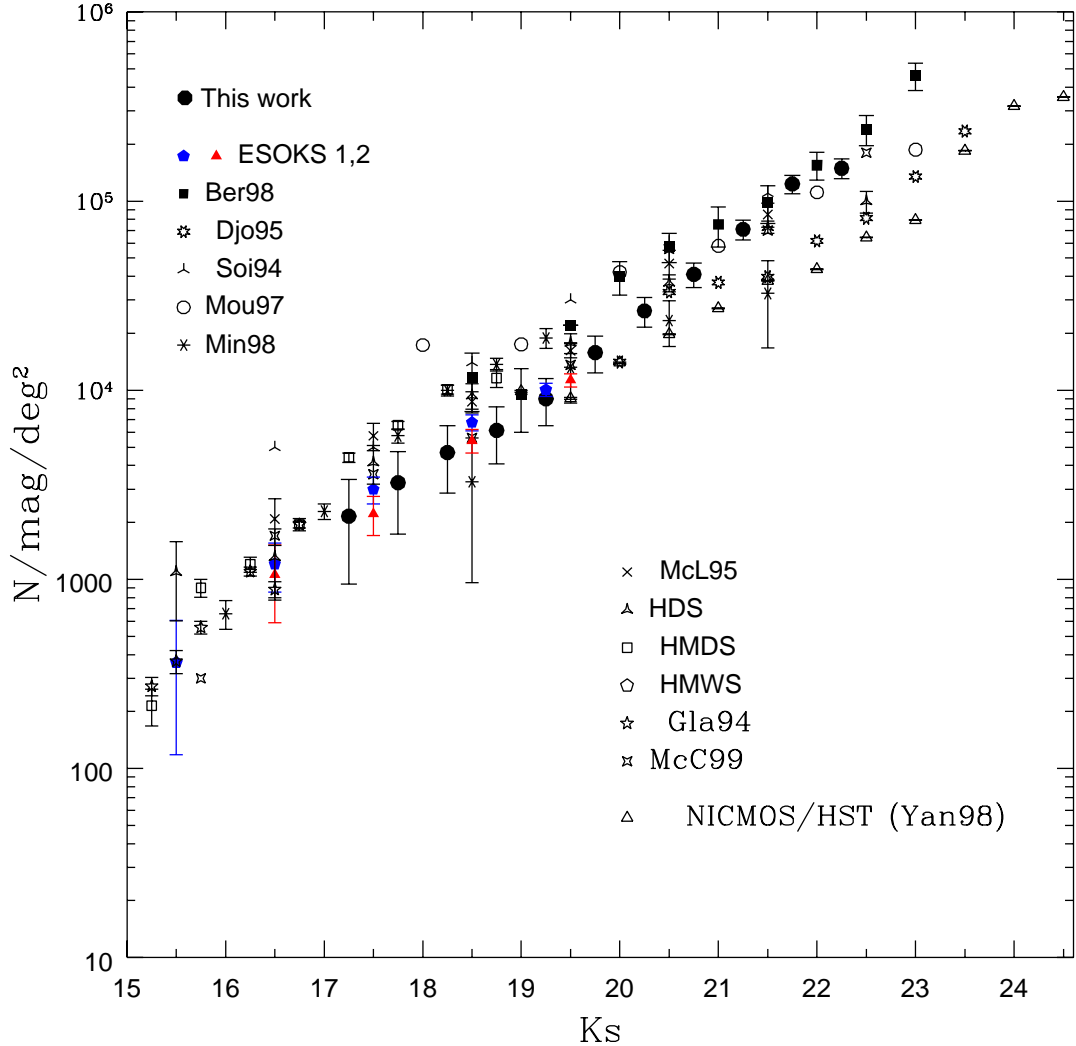,height=150mm}}
 \caption{The figure compares the Ks-band counts obtained in this 
work with those in the literature: 
  Djorgovski et al. (1995, Djo95), Bershady et al. (1998, Ber98) 
(obtained at the Keck telescope), Gardner et al. (1993, HMWS, HMDS, HDS), 
Glazebrook et al. (1994, Gla94), Soifer et al. (1994, Soi94), 
McLeod et al. (1995, McL95), Moustakas et al. (1997, Mou97),
Saracco et al. (1997, ESOKS1, 2), Minezaki et al. (1998, Min98) and
McCracken et al. (1999, McC99). 
NICMOS/HST H-band counts of Yan et al. (1998, Yan98) are also shown.
The slope of our counts ($\sim$0.38) agree very well with those obtained
by McL95, Djo95 and Ber98 while it is significantly steeper than
the slope obtained by Gardner et al. (1993) and by Mou97.}
\end{figure*}

\begin{table}
\caption{ Differential number counts in the Ks-band (upper panel) 
and in the J-band (lower panel) derived from the samples including
sources having S/N$\ge$5.
Errors $\sigma_N$ take into account the Poissonian error on raw counts ($n_r$),
the uncertainties in the incompleteness correction factor $c$ and the 
fluctuations due to clustering (see text). The amplitude $A_\omega$ of the
angular correlation function has been obtained using the scale relation
log($A_\omega$)$\sim-0.3$mag+cost.}  
\begin{center}
 \begin{tabular}{crcrrl}
\hline
\hline
     Ks &  $n_r$  & $c$ & $N/mag/deg^2$ &$\sigma_N$&  $A_\omega(\theta'')$  \\
\hline
    15-17&    4&  1.0  &      360 &	 *** &*** \\	
    17.25&    6&  1.0  &      2160 &    1215 &8.2993\\	
    17.75&    9&  1.0  &       3240 &    1510&5.8755\\
    18.25&  13 &  1.0  &      4680 &   1825  &4.1595\\
    18.75&   17&  1.0  &      6120  &   2048 &2.9447\\
    19.25&   25&  1.0  &      9000  &   2508 &2.0847\\
    19.75&   44&  1.0  &     15840  &   3520 &1.4758\\
    20.25&   73&  1.0  &     26280  &   4743 &1.0448\\
    20.75&  114&  1.0  &     41040  &   6106 &0.7397\\
    21.25&  197&  1.0  &    70920   &  8550  &0.5237\\
    21.75&  263&  1.3  &    123100 &  13898  &0.3707\\
    22.25&  109&  3.7  &    145200 &  17987  &0.2624\\
\hline
\hline
        J &  $n_r$  & $c$ & $N/mag/deg^2$ & $\sigma_N$& $A_\omega(\theta'')$\\
\hline
    16-18&    4&  1.0&       360 &	***&	***\\
    18-19&   12&  1.0&	     2160&	1215&  8.2993\\	
    19.25&   10&  1.0&       3600&    1634&   5.8755\\
    19.75&   15&  1.0&       5400&    2033&   4.1595\\
    20.25&   18&  1.0&       6480&    2137&   2.9447\\
    20.75&   33&  1.0&      11880&    3097&   2.0847\\
    21.25&   64&  1.0&      23040&    4738&   1.4758\\
    21.75&   82&  1.0&      29520&    5203&   1.0448\\
    22.25&  153&  1.0&      55080&    7770&   0.7397\\
    22.75&  253&  1.0&      91080&   10548&   0.5237\\
    23.25&  417&  1.2&     180144&   21349&   0.3707\\
    23.75&  224&  2.8&     225800&   23828&   0.2624\\
\hline
\hline
\end{tabular} 
\end{center}
\end{table}

The raw counts $n_r$, the completeness correction factors $c$, 
the counts per square degree corrected for incompleteness $N$ and 
their errors $\sigma_N$ are reported in Tab.2.
The number uncertainty $\sigma_N$ has been derived by quadratically summing up 
the Poissonian error $\sigma_{n_r}=\sqrt{n_r}$ of raw counts, the uncertainty 
in the completeness factor $\sigma_c=(N/n_r)\sigma_{n_r}$ and the contribution 
$\sigma_\omega$ due to clustering fluctuations.
For an angular correlation function with a power law form
$\omega(\theta)=A_\omega\theta^{-(\gamma-1)}$ ($\gamma=-1.8$)
and a circular window function
of radius $\theta_0$, the expected fluctuations are

\begin{equation}
\sigma_\omega\sim\omega(\theta)^{1/2}\bar N
\end{equation}
We have assumed that the amplitude $A_\omega$ evolves with magnitude
following the relation log$A_\omega\sim-0.3\times$mag (Brainerd et al. 1994;
Roche et al. 1996)
and that $\theta_0\sim150''$, which is the  radius of a circle having
an area corresponding to our fields.
In Tab.2 the derived amplitude $A_\omega$ is also reported for each
magnitude bin.
The values of the amplitudes are consistent with
the results in the literature based on infrared selected samples 
(Roche et al. 1998; Carlberg et al. 1997.
It is worth noting that while at bright magnitudes (Ks$<19$ and J$<20$)
the poissonian error is comparable to the uncertainties introduced by
clustering fluctuations ($\sigma_{n_r}\sim\sigma_{\omega}$) 
at magnitudes Ks$\sim$22 and J$\sim$23 
$\sigma_{\omega}=2\times\sigma_{n_r}$ and,
in the faintest magnitude bins, the single contributions to the total
uncertainty $\sigma_{N}$  are among them as $\sigma_{n_r}:\sigma_{c}:\sigma_{\omega}=1:3.8:2.7$.

The total number of sources  in Tab.2 are 874 for the Ks-band
and 1285 for the J-band and represent the largest samples currently
available at these depths.
\begin{figure}
\centerline{\psfig{figure=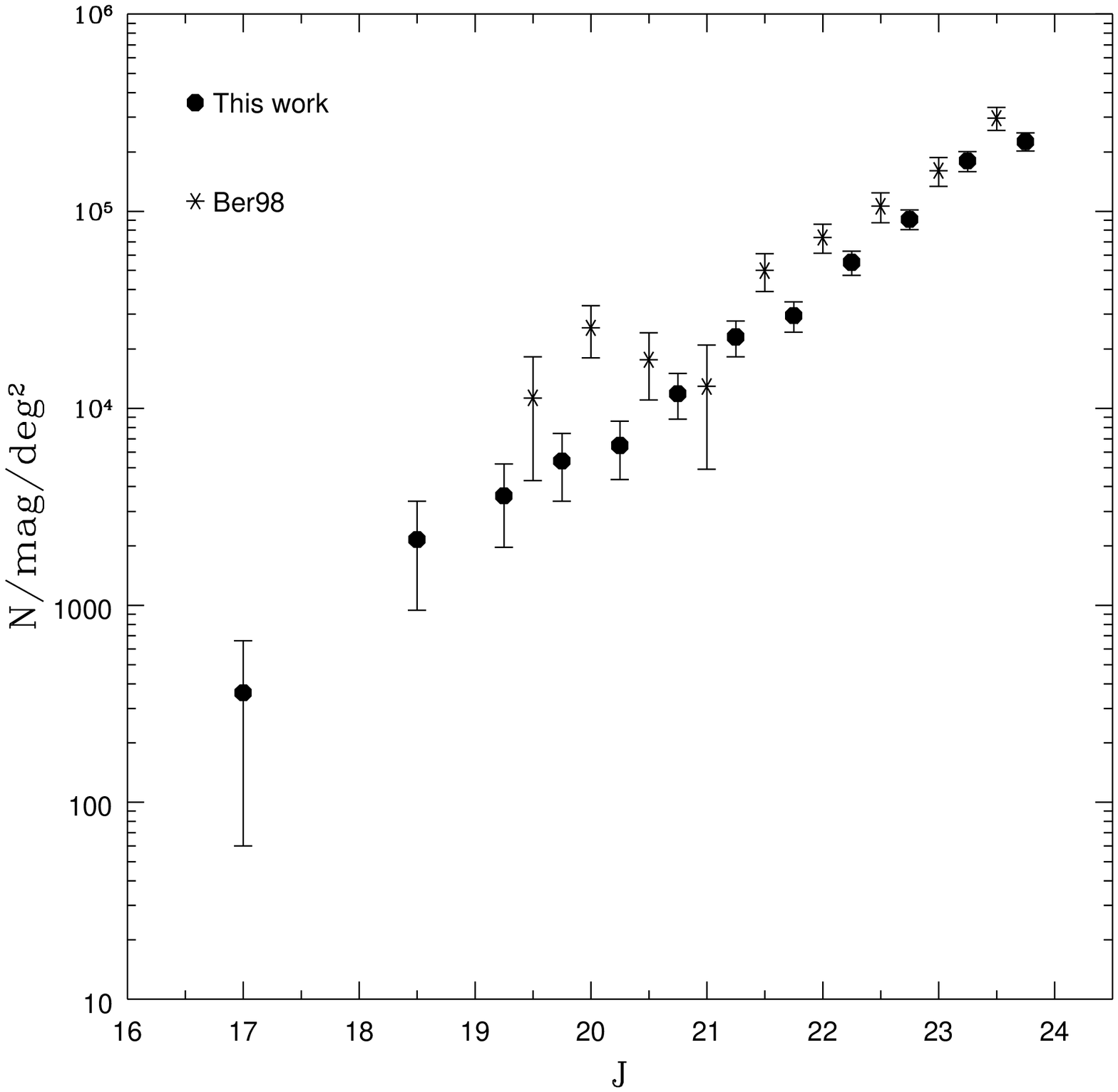,height=90mm}}
 \caption{The J-band galaxy counts obtained in this work are
compared with those previously obtained by Bershady et al. (1998, Ber98).
The J-band counts continue to rise with a power law slope of  $\sim$0.36 
 showing  no signs of flattening down to $J=24.0$.}
\end{figure}
 
Fig. 3  shows the Ks-band galaxy 
counts  derived here compared with those in 
the literature{\footnote{A recent paper by Ferreras et al. (1998) presents
galaxy counts down to K=19 but they do not provide sufficient data
to derive (and thus reproduce here) their counts.}}.
Our counts follow a d$logN/dK$ relation with a slope of 0.38 in the
magnitude range $17<K<22.5$ and do not show  evidence of any turnover or
 flattening down to the limits of the survey.

In Fig. 4 our J-band galaxy counts are compared with those of Bershady et al.
(1998) which are the only J-band data available in the literature.
For the J number counts
we obtain a slope  $\sim$0.36 in the magnitude
interval $18<J<24$.
As for the Ks trend, J counts do not show any sign of  
flattening down to the limits of the data.

\subsection{Galaxy Colors}
The 2.5 arcsec aperture J-K color of galaxies in our sample was obtained
by running SExtractor in the so-called {\em  double-image mode}:
J magnitudes for the Ks  selected sample have been derived
using the Ks frame as  reference image for detection and the J image 
for measurements only and {\it vice versa} for the J selected sample.
This method forces the program to carry out a net flux measurement 
within the aperture centered on the position of the source detected 
in the reference frame.

In order to avoid unreliable measurements, that is net flux simply due
to local background fluctuations, we have considered as genuine
estimates those signals exceeding $1.5\sigma$ above the background
within the 2.5 arcsec diameter aperture (Ks$_{lim}=23.05$, J$_{lim}=24.8$).  Fainter signals have been
considered unreliable and they have been replaced by a $1.5\sigma$
upper limit.  In Tab.3 the number of galaxies, the median J-Ks color,
the standard deviation and the number of lower and upper limits to
J-Ks in each magnitude bin are reported for the Ks and J sample,
respectively.

In Figs. 5  and 6  color-magnitude diagrams of the whole 
Ks and J galaxy samples are shown together with the median locus.
The error bars are the standard deviation from the mean of the values
within each bin.
\begin{figure}
\centerline{\psfig{figure=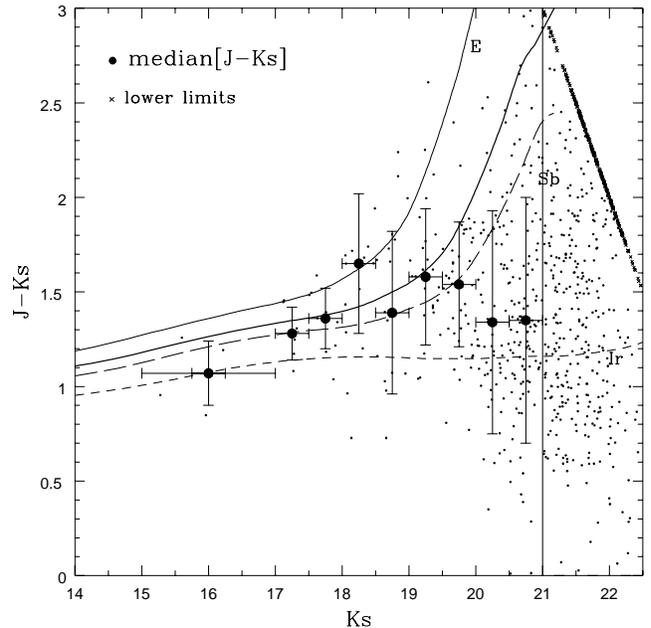,height=90mm}}
 \caption{Color-magnitude diagram of our Ks galaxy sample.
Large filled points represent the median J-Ks color. 
Vertical error bars are the standard 
deviation from the mean. 
The bin widths (usually 0.5 mag) are represented by the 
horizontal bars. 
Small crosses mark undetected J galaxies, i.e. the J-Ks lower limits.
The vertical straight line represents the completeness J-band detection limit 
for the Ks sample.
The thick solid  curve is the expected apparent color of a local galaxy 
mix (see \S 6.2 and Tab. 4) evolved back
to high redshift ($z>10$) according to the models of Buzzoni (1998).
The expected J-K color for a pure galaxy population of E (thin solid line), 
Sb (long-dashed line) and Im (short-dashed line) types is also
reported  as labeled  in the figure. Models assume a
($H_o, q_o = 50, 0.5$) cosmology.}
\end{figure}

\begin{figure}
\centerline{\psfig{figure=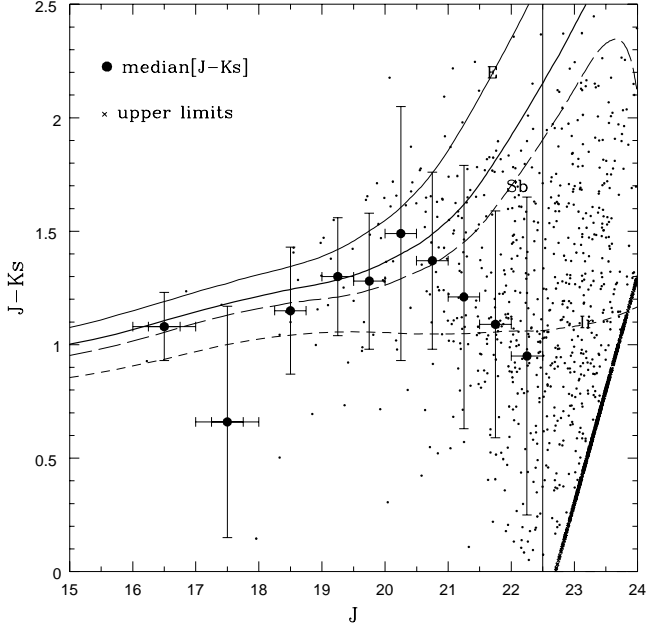,height=90mm}}
 \caption{Same as Fig. 5 but for the J-selected galaxies.
The vertical solid line indicates the detection limit of the J sample
in the Ks frame.}
\end{figure}

The most notable feature in both figures is a break in the color trend. 
The intervening redder J-Ks with increasing magnitude, 
as induced by the k-correction, turns in facts to bluer values for the 
galaxies fainter than $Ks \sim 19$ or $J \sim 20$.
We will discuss this important issue in \S 6.

\section{Galaxy Size and Number Counts}
\subsection{Compactness Index}
A reasonable measure of the apparent size of distant galaxies,
represents a critical step in the analysis of faint galaxy samples
(Petrosian 1998) and hence in the application of cosmological
tests. Given the cosmological dimming of the apparent surface
brightness (SB), galaxy isophotal radius would be a poor measure of
apparent size, especially for the faintest objects in a deep survey.
A fixed isophotal threshold would not necessarily span a
similar absolute size of galaxies at different redshifts.  Special
care has therefore to be taken in estimating the galaxy
metric size in a way which is independent on the redshift and source profile
and takes into account the photometric bias (Djorgovski \& Spinrad
1981; Kron 1995; Petrosian 1998).

For our samples, we adopted the metric size function $\eta(\theta)$ 
first introduced by Petrosian (1976) defined as
\begin{equation}
\eta(\theta)={{1~d~ln~l(\theta)}\over{2~d~ln(\theta)}}
\end{equation}
(Kron 1995) where $l(\theta)$ is the growth curve.
This function has the property
\begin{equation}
\eta(\theta)={{I(\theta)}\over{\langle I\rangle}_\theta}
\end{equation}
where  $I(\theta)$ is the surface brightness at radius
$\theta$ and ${\langle I\rangle}_\theta$ is the mean surface brightness 
within $\theta$.

The function $\eta(\theta)$ has been obtained for each galaxy by
constructing its own intensity profile.  The intensity profiles have
been derived from the J and Ks frames through equi-spaced (0.146
arcsec) multi-aperture photometry after a $2\times2$ re-sampling of
the images.  Following Bershady et al. (1998), we define the galaxy 
angular size the value of $\theta_{\eta}$ such that $\eta(\theta_{\eta})=0.5$
as obtained by a spline fit interpolation across each object.  The
median $\theta_{0.5}$ values are 0.73 arcsec and 0.79 arcsec for the J
and Ks samples, respectively.  
For reference,  the measured angular size of point sources
in our images is $\theta_{0.5}=0.67$ arcsec.  
In order to simplify comparison of
these angular sizes with those in the literature we have also measured
for each galaxy the effective radius $\theta_{eff}$.  The result is that
$\theta_{eff}\simeq0.6\cdot \theta_{0.5}$ for our samples,  which
agrees with the effective radius measured by Yan et al. (1998).

According to the fiducial galaxy size, we were also able to derive a 
compactness parameter via the luminosity concentration index $C_\eta$

\begin{equation}
C_\eta={{F(<\theta_{0.5})}\over{F(<1.5\theta_{0.5})}}
\end{equation}
that is the ratio between the flux within the radius $\theta_{0.5}$
and that within $1.5\theta_{0.5}$.  This dimensionless parameter
varies in the range $0<C_\eta<1$  increasing in the most compact
objects and vanishing in amorphous  galaxies. 
The estimated values of $C_\eta$ for  simulated spiral 
and  elliptical profiles (convolved with the observed PSF) 
are 0.69 and 0.74 respectively, to be compared with $C_\eta=0.9$ of the PSF.
 Note that, by definition,
$C_\eta$ is free of any adopted surface brightness threshold in the
photometry, and is therefore {\it not} affected by galaxy redshift
as for instance the central concentration index C defined by Abraham
et al. (1994; 1996) and discussed by Brinchmann et al. (1998).  
The concentration index $C_\eta$ and the apparent radius $\theta_{0.5}$
for the whole Ks and J samples are given in Fig. 7 and 8,
respectively, together with their median locus.  
In the figures, error bars are the standard deviation from the mean 
within each bin.  
The median values $med[\theta_{0.5}]$ and $med[C_\eta]$ and the relative
standard deviations are listed in Tab. 3.

\begin{figure}
\centerline{\psfig{figure=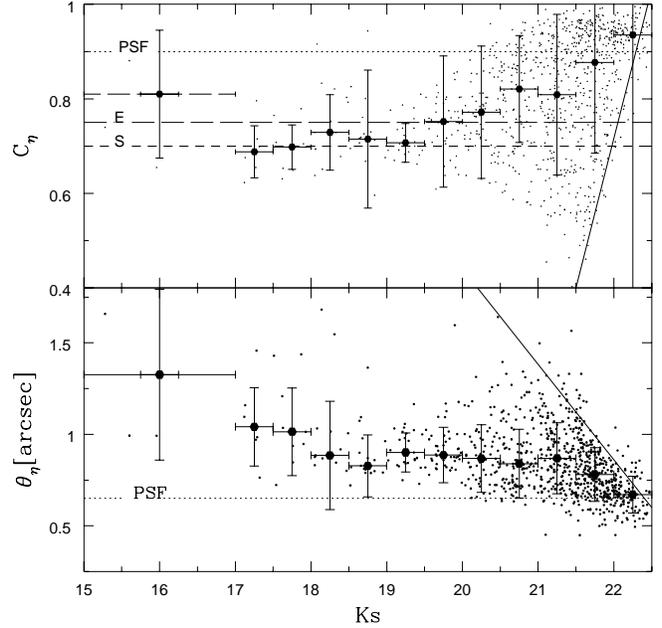,height=90mm}}
 \caption{Compactness index $C_\eta$ (upper panel) 
and apparent size $\theta_{0.5}$ (lower panel) as a function of Ks magnitude.
The filled (red) circles are the median values in a 0.5  mag bin width.
Vertical error bars are the standard deviation in each magnitude bin
while horizontal bars represent the bin width.
The dotted, the dashed and the long-dashed lines in the upper panel
indicate the values of $C_\eta$ for PSF, Spiral and Elliptical profiles
respectively (see \S 5.2).
The solid line in the lower panel shows the magnitude enclosed in 
an aperture of radius $\theta$ and a uniform surface brightness 
Ks=22.76 mag/arcsec$^2$ while the dotted line represent the characteristic
size of a point source in our images.
}
\end{figure}

\begin{figure}
\centerline{\psfig{figure=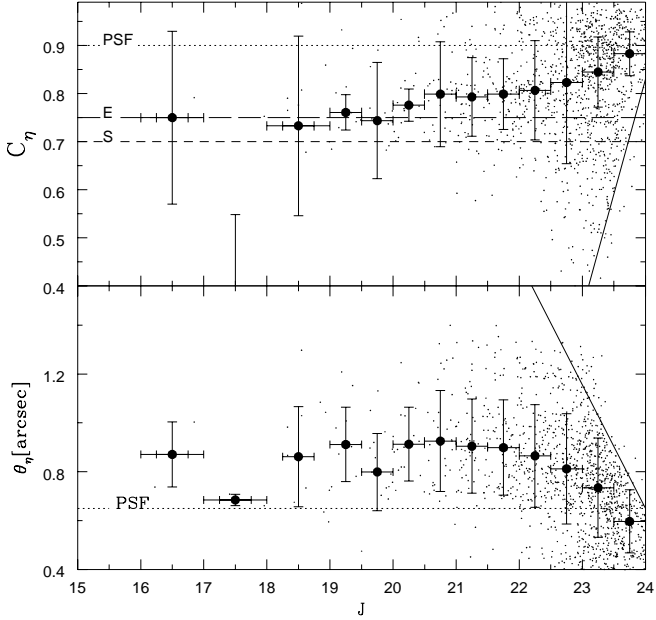,height=90mm}}
 \caption{Same as in Fig. 7 but for the J sample.
The solid line in the lower panel shows the magnitude enclosed in 
an aperture of radius $\theta$ and a uniform surface brightness of
J=23.88 mag/arcsec$^2$.}
\end{figure} 

The compactness index seems to be constant down to
Ks$\sim19$ and J$\sim20$ whereas going to fainter magnitudes 
it is evident that the galaxies become  systematically  more compact.
This trend is clearly present in both the samples as shown by the 
increase of the median values of $C_\eta$ with increasing apparent magnitude.
It is worth noting that the increasing compactness of galaxies
occurs in the same magnitude range where galaxies also become  bluer.
Moreover, this trend seems to be matched also by a decrease of the 
apparent size of galaxies, a trend which is more evident in the Ks 
sample (Fig. 7) and which was found also by Yan et al. (1998).
However both apparent size and color depend on redshift and while
the former change less than 20$\%$ beyond $z=0.7$ the latter continues
to change substantially.
 We would like to stress that, while the apparent size 
is a redshift dependent parameter and hence its systematic decrease to
 fainter magnitudes could be attributed to a systematic increase 
of redshift,
{\em the compactness index defined by (4) is redshift independent and its
trend with apparent magnitude thus represents a real physical behavior of 
faint galaxies}. 

\begin{table*}
\caption{Median (J-Ks) color, size and compactness index
of the galaxy sample. Errors are the standard deviation in
each bin. The number of lower and upper limits (L.L., U.L.) 
represents, in the upper panel, the number of galaxies in each magnitude bin
belonging to the Ks sample undetected in the J frame and, in the lower panel, 
the J-band selected galaxies  undetected in the Ks image.
Lower and upper limits to the J-K color are marked with $^\uparrow$
and $^\downarrow$ respectively.}
\begin{center}
 \begin{tabular}{crcrrrlrl}
\hline
\hline
 Ks&  Gal& med[J-K]&$\sigma$ &L.L.&med[$\theta_{0.5}$]&$\sigma$& 
med[$C_{\eta}$]&$\sigma$\\
\hline
 16.00&   4&   1.07&   0.2&    0&   1.33&   0.5&      0.81&  0.1\\
 17.25&   6&   1.28&   0.1&    0&   1.04&   0.2&      0.69& 0.1\\
 17.75&   9&   1.36&   0.2&    0&   1.01&   0.2&      0.70&  0.05\\
 18.25&  13&   1.65&   0.4&    0&   0.88&   0.3&      0.73& 0.1\\
 18.75&  17&   1.39&   0.4&    0&   0.83&   0.2&      0.71&  0.1\\
 19.25&  25&   1.58&   0.4&    0&   0.90&   0.1&      0.71&  0.04\\
 19.75&  44&   1.54&   0.3&    0&   0.89&   0.1&      0.75&  0.1\\
 20.25&  73&   1.34&   0.6&    1&   0.87&   0.2&      0.77&  0.1\\
 20.75& 114&   1.35&   0.7&    0&   0.84&   0.2&      0.82& 0.1\\
 21.25& 197&   1.65$^\downarrow$&   0.8&   24&   0.87&   0.2&      0.81&  0.2\\
 21.75& 263&   2.16$^\downarrow$&   0.7&  109&   0.78&   0.1&      0.88&  0.2\\
 22.25& 109&   1.84$^\downarrow$&   0.7&   50&   0.67&   0.1&      0.93&  0.7\\
\hline
\hline

J&  Gal& med[J-K]&$\sigma$ &U.L.&med[$\theta_{0.5}$]&$\sigma$& 
med[$C_{\eta}$]&$\sigma$\\
\hline
 16.50&   2&  1.08&   0.1&    0&   0.87&   0.1&      0.40&  0.4\\
 17.50&   2&  0.66&   0.5&    0&   0.68&   0.02&     0.38&  0.4\\
 18.50&  12&  1.15&   0.3&    0&   0.86&   0.2&      0.73&  0.4\\
 19.25&  10&  1.28&   0.3&    0&   0.95&   0.2&      0.78&  0.1\\
 19.75&  15&  1.28&   0.3&    0&   0.80&   0.2&      0.74&  0.3\\
 20.25&  18&  1.49&   0.6&    0&   0.91&   0.1&      0.78&  0.07\\
 20.75&  33&  1.37&   0.4&    0&   0.93&   0.2&      0.80&  0.2\\
 21.25&  64&  1.21&   0.6&    1&   0.90&   0.2&      0.79&  0.2\\
 21.75&  82&  1.09&   0.5&    0&   0.90&   0.2&      0.80&  0.2\\
 22.25& 153&  0.95$^\uparrow$&   0.6&   13&   0.86&   0.2&      0.81&  1.1\\
 22.75& 253&  0.63$^\uparrow$&   0.6&   85&   0.81&   0.2&      0.82&  0.4\\
 23.25& 417&  0.86$^\uparrow$&   0.4&  239&   0.73&   0.2&      0.84&  0.2\\
 23.75& 224&  1.22$^\uparrow$&   0.3&  150&   0.60&   0.1&      0.88&  0.1\\
\hline
\hline
\end{tabular}
\end{center}
\end{table*}

\subsection{Number Counts vs Size}
Following Bershady et al. (1998), we divided the J and Ks galaxy 
population in two
sub-samples of small ($s$) and large ($l$) objects relative to 
the median $\theta_{0.5}$ value
(i.e. med[$\theta^J_{0.5}$]=0.73 arcsec, 
med[$\theta^{Ks}_{0.5}$]=0.79 arcsec).

A suitable completeness estimate is a critical step in this case.  For
a given total magnitude the mean surface brightness of a source
depends on its size so that larger galaxies tend to be less visible
than smaller ones.  Fortunately, the simulation procedure we devised
in \S 5.1 allowed us to completely control this selection effect and
derive a suitable completeness correction.

The differential galaxy counts for the $s$ and $l$ galaxies are shown
in Fig. 9 which clearly reveals the different behavior followed by the
$s$ compared to the $l$ sources: small sources show a lower surface
number density at bright magnitudes, and a much steeper counts slope
over the whole magnitude range.  The slopes we derived for the $s$
samples are essentially {\em Euclidean}.
In particular we obtained $\gamma^s_{Ks}\sim0.6$ and 
$\gamma^s_{J}\sim0.56$ to be
compared with $\gamma^l_{Ks}\sim0.36$ and $\gamma^l_{J}\sim0.38$ for
the $l$ sample.  Supporting Bershady's et al. (1998) conclusions
therefore, small sources dominate the number counts at faint
magnitudes (Ks$>$21.5, J$>23$) where their number density is more than
twice the number density of large sources.

\begin{figure}
\vspace {-2truecm}
\centerline{\psfig{figure=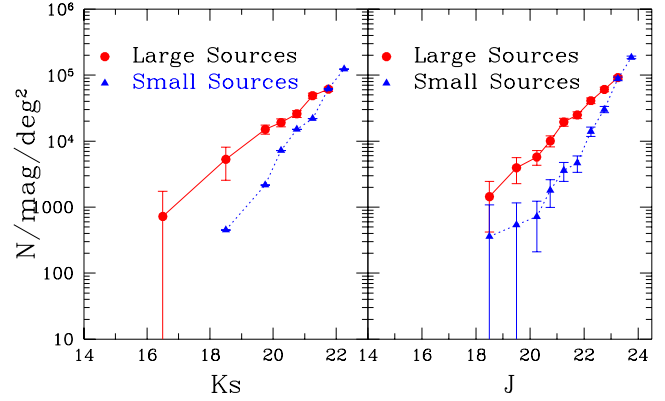,height=90mm}}
 \caption{Ks-band (left) and J-band (right) differential number counts 
for small and large galaxies: small galaxies are those having 
$\theta^K_{0.5}\le0.79$ arcsec and $\theta^J_{0.5}\le0.73$ arcsec.
The slope of the counts are $\gamma^s_{Ks}\sim0.6$
and $\gamma^s_{J}\sim0.56$ for small sources and $\gamma^l_{Ks}\sim0.36$ and 
$\gamma^l_{J}\sim0.38$ for large sources.}
\end{figure}

In Fig. 10 and 11 the J-Ks color distributions of $s$ and $l$
galaxies are shown for the Ks and J sample respectively.
The shaded histograms in each figure represent the distribution
of the J-Ks lower limits (for the Ks selected sample) and upper limits
(for the J selected sample).

Large sources tend to have a sharper distribution and a redder median
J-Ks color compared to $s$ sources.
Moreover, the latter   are skewed toward bluer colors.  
Such differences between the J-Ks color distributions of $s$ and $l$ sources
 are highly significant for both samples.
The two distributions have been first compared by performing a K-S test 
on the complete color samples, that is samples limited to 
Ks$\le21$ and J$\le22$, in order to avoid uncertainties introduced 
by the presence of censored data.
In addition, we have extended the comparison to the whole sample by making use 
of  ``survival analysis'' (Avni et al. 1980; Isobe et al. 1986; Feigelson 
and Nelson 1985).
In particular we have compared the two sample distributions with the  
generalized Wilcoxon test, which is able to treat data including
lower and upper limits equally and obtaining a high significance (10$^{-3}$).
 
In Tab. 4 we report the median J-Ks in 1 mag bin width 
for $s$ and $l$ sources of the Ks and J sample while Fig. 12 shows
the color-magnitude diagrams of $s$ and $l$ sources.
It can be seen the higher fraction of blue galaxies in  the sample
of $s$ sources at faint magnitudes with respect to $l$ sources as confirmed
by the values reported in Tab. 4.

\begin{figure}
\centerline{\psfig{figure=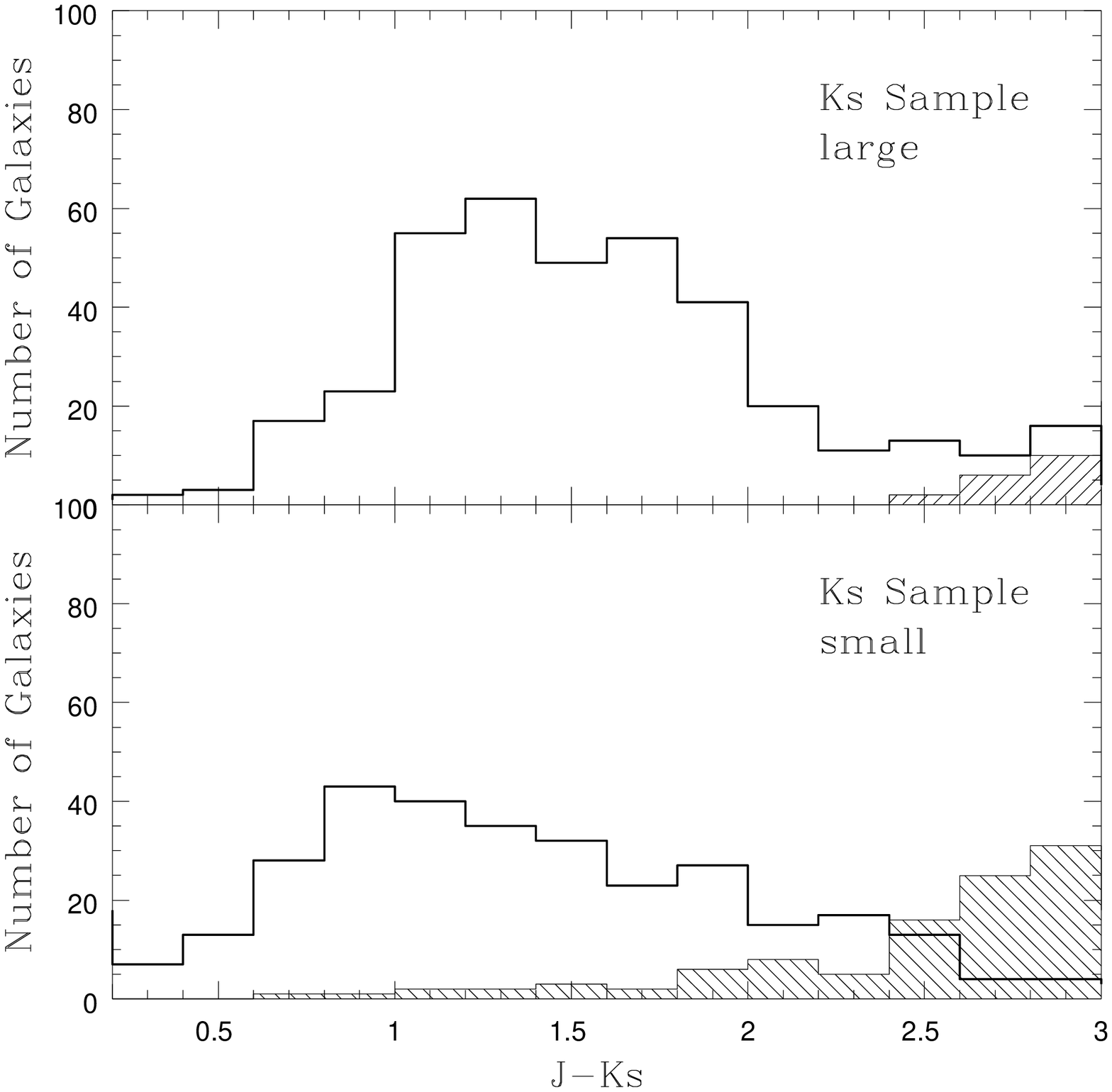,height=90mm}}
 \caption{J-Ks color distributions (thik histograms) for $l$ 
(upper panel) and $s$ sources (lower panel) belonging to the Ks sample. 
The shaded histograms represent the distribution of the J-Ks
lower limits i.e. galaxies redder than $J_{lim}-Ks$ (J$_{lim}=24.8$, \S 4.2).}
\end{figure}

\begin{figure}
\centerline{\psfig{figure=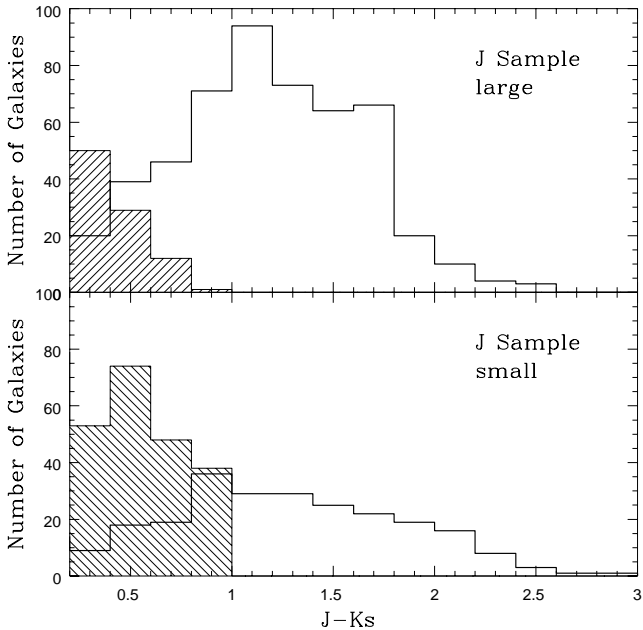,height=90mm}}
 \caption{J-Ks color distribution for $l$ (upper panel) and
$s$ sources (lower panel) belonging to the J sample. 
The shaded histograms represent the distribution of the J-Ks
upper limits i.e. galaxies bluer than $J-Ks_{lim}$ (Ks$_{lim}=23.05$, \S 4.2).}
\end{figure}

\begin{figure}
\centerline{\psfig{figure=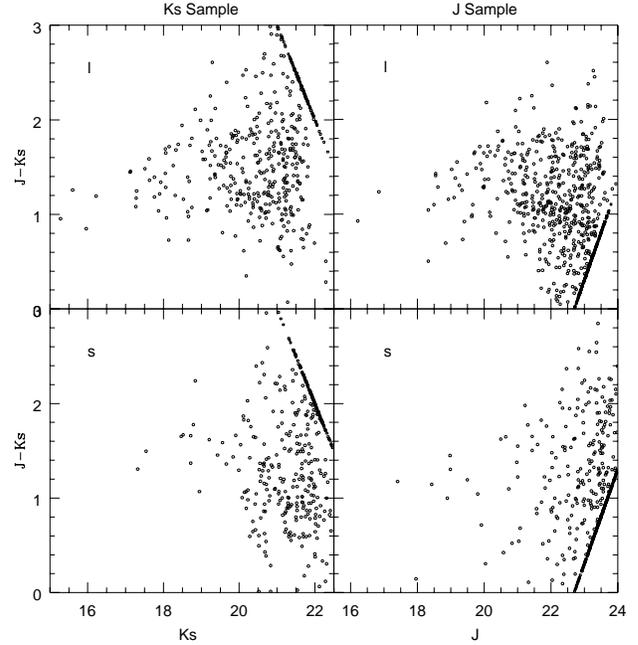,height=90mm}}
 \caption{Color-magnitude diagram of $s$ (lower panel)
and $l$ (upper panel) sources for Ks sample (left) and J
sample (right).}
\end{figure}

\begin{table}
\caption{Observed median (J-Ks) color as a function
of apparent magnitude for {\em s} and {\em l} sources in the Ks and J 
samples.
Lower and upper limits to the J-K color are marked with $^\uparrow$
and $^\downarrow$ respectively.}
\begin{center}
 \begin{tabular}{clllllll}
\hline
\hline
Class&17.5&18.5&19.5&20.5&21.5&22.5&23.5\\
\hline
$s_K$&1.40&1.62&1.52&1.17&1.82$^\downarrow$&1.48$^\downarrow$& ---\\
$l_K$&1.34&1.48&1.54&1.45&2.01$^\downarrow$&1.83$^\downarrow$& ---\\
\hline
$s_J$&--- &0.66&0.69&0.72&0.70&0.67$^\uparrow$&0.57$^\uparrow$\\
$l_J$&--- &1.13&1.34&1.40&1.12&1.02$^\uparrow$&0.78$^\uparrow$\\
\hline
\hline
\end{tabular}
\end{center}
\end{table}

\section{Discussion}
\subsection{Galaxy Counts}
In Fig. 3 we have compared our Ks-band galaxy counts with those
derived by other authors.  The remarkable scatter between the
different surveys is probably due to a combination of various factors:
slightly different filters (K, K', Ks), differences in the treatment
of star contamination and in the magnitude estimate (aperture or
isophotal or pseudo-total), large scale fluctuations, field selection
criteria and, finally, possible systematics in the photometric
calibrations.  The largest deviation occurs between the Soifer et
al. (1994) counts, defining the highest surface density at magnitudes
K$>19$, and those of Djorgovski et al. (1995) and Minezaki et
al. (1998) which define the lowest surface density of objects in this
magnitude range.  The amplitude of our counts at magnitude brighter
than Ks=20 agrees with that derived by Glazebrook et al. (1994),
Saracco et al. (1997) and with that obtained by Minezaki et al. (1998)
in the bright survey, while it is significantly lower than the
amplitude of HMDS and HDS (Gardner et al. 1993) and of McLeod et
al. (1995).  At fainter magnitudes (Ks$\ge20$) we obtain a surface
density of objects consistent with that of Bershady et al. (1998) and
McLeod et al. (1995) but significantly higher than that of HDS
(Gardner et al. 1993; Cowie et al.  1994) and Djorgovski et
al. (1995).  It is worth noting that the area covered by these
surveys, and consequently the number of sources, is at most one fifth
of that surveyed in this work.  This obviously amplifies the
scattering due to large-scale fluctuations.  However, rather than the
amplitude, the most striking feature of the plot is the slope of the
number counts.  Contrary to Gardner et al. (1993) and Moustakas et
al. (1997) who obtain a slope in the range 0.23-0.26 for $18<K<22$, we
obtain a significantly steeper slope of 0.38.  This result confirms
the previous work by McLeod et al. (1995), Soifer et al. (1994) and
Saracco et al. (1997) at K$<21$ and Djorgovski et al. (1995),
Bershady et al.  (1998) and McCracken et al. (1999) down to K=23-24.
Moreover, the slope of our K-band counts, and the lack of any
flattening at its faint end, are both independently confirmed also by
the J-band counts which indicate a slope of $\sim0.36$ in agreement with
the result of Bershady et al.  The overall conclusion is therefore
that {\em galaxy counts in the near IR continue to rise with no
evidence of a turnover or a flattening down to the current magnitude
limit of the observations.}

\subsection{Colors and Galaxy Counts}
Which galaxies are responsible for the continuous increase of IR counts
at very faint magnitudes ?
The two main observational results  indicate that
{\it i)} at Ks$>19$ and J$>20.5$ the median J-Ks color of the galaxy 
population turns
bluer, as shown in Fig. 5 and 6 and summarized in Tab. 3;
{\it ii)} the same surface density of objects at Ks$\sim$22 is 
reached at J$\sim23$, 
which implies that the bulk of the population of galaxies at these
apparent magnitudes should have a mean color J-Ks$\sim 1$, in agreement with
the observed J-Ks color trend.

In Fig. 13 the expected J-K color of galaxies for different morphological
types and that expected  for the local galaxy mix
 are shown as a function of redshift for evolving and non evolving galaxies. 
Model predictions are based on Buzzoni's (1989, 1995) population 
synthesis code. 
Three sets of model galaxies have been computed to ease comparison with 
the observations. 
The evolution of ellipticals (E) is accompanied with that of $Sb$ 
spirals and $Im$ irregulars.
As in the standard scenario, ellipticals are assumed to be suitably 
described by a single burst stellar population while irregulars follow a 
flat star formation rate at every age. 
The intermediate case of $Sb$ spirals , on the contrary, takes into account
a declining star formation rate in the disk ($SFR \propto t^{-0.5}$) 
coupled with a spheroidal component (bulge+halo) as in the ellipticals 
(see Buzzoni 1999 for more details on the model templates). 
From the figure, galaxies with J-Ks$\sim1$ are compatible with redshifts 
$z \sim 0.2-0.3$, and a value of  J-K$<1.5$ seems to be   consistent 
with a reshift no larger than $z\sim1$ as shown by the model representing
the mean color weighted by the local galaxy mix (Tot).
However it should be mentioned that IR colors alone are not 
sufficient to strongly constraint redshifts. 

One major advantage of the $J$ and $K_s$ infrared data is 
that we are probing the flatter and relatively shallower region of the
rest-frame galaxy spectral energy distribution.
The only significant flux discontinuity, coming from the Balmer break, 
only enters the $J$ band well beyond $z \sim 2$. In any case, 
the net effect of redshift on the apparent infrared colors is a 
smooth reddening with increasing distance resulting in  high-$z$ galaxies 
somewhat resembling local $M$-type stars in the Galactic field.
More generally, therefore, up to $z \sim 3$, a sequence in apparent $J-K$ 
color can also be regarded as a monotonic sequence in distance for 
galaxies of the same morphological type.
In figures 5 and 6, the theoretical J-K colors for the three different
morphological types are compared  with those observed as a function 
of the apparent J and Ks magnitudes.
In addition, we also show the expected colors for the local galaxy 
mix as a whole derived by taking into account the present-day luminosity 
function as in Marzke et al. (1998). 
Operationally, the $B$ Schechter parameters for the different morphological 
types (cf. Tab. 1 in Marzke {\it et al.} 1998) have been combined with the 
theoretical $B-J$ and $B-K$ colors to obtain the mean $J$ and $K$ 
luminosity contribution from early, late and irregular galaxies.
For each galaxy subsample ($i$), this is simply assumed to be proportional to 
$\int L_i\phi_i(L)$ where $\phi_i(L)$ is the appropriate Schechter 
function so that
\begin{equation}
{{{L_i}}\over {L_{TOT}}} = 
{{\Gamma(2+\alpha_i)\ \phi^*_i\ L^*_i} \over 
{\sum \Gamma(2+\alpha_i)\ \phi^*_i\ L^*_i}}
\end{equation}
summing up over the running index $i$ for ellipticals, spirals and irregulars.
For each galaxy template and for the local galaxy mix we finally computed 
the apparent $J-K$ color as well as the apparent magnitude of a reference 
$M^*$ galaxy in a $(H_0, q_0) = (50, 0.5)$ cosmology.
The resulting luminosity partition in the different (rest-frame) photometric 
bands at $z \to  0$ is summarized in Tab. 5 (note of course that the Marzke 
{\it et al.} relevant values for $M^*$ and $\phi^*$ have been rescaled 
in the Table according to $H_0=50$ Km s$^{-1}$ Mpc$^{-1}$).

\begin{table*}
\caption{Type-dependent LFs,  optical-near-IR colors and morphological mix
used in the models}
\begin{center}

 \begin{tabular}{cccccccccccc}

\hline \hline Type & $\alpha$& $\phi_*$& $M^*_B$& B-J&J-K &
$M^*_J$&$M^*_K$& $f_B$&$f_J$&$f_K$\\ & & $10^{-4}$ Mpc$^{-3}$& & & & &
& & &\\ \hline E & -1.00& 5.5& -20.88& 3.19& 1.03& -24.07& -25.10&
29$\%$& 44$\%$& 46$\%$\\

Sb& -1.11&  10.0&  -20.94&  2.58&  0.94& -23.52& -24.46&    61$\%$&   51$\%$& 
 49$\%$\\

Irr&  -1.81&   0.2& -21.29&  2.10&  0.85& -23.39& -24.24&    10$\%$&    5$\%$&
 5$\%$\\

\hline
\hline
\end{tabular} 
\end{center}
\end{table*}
Since model refers to $L^*$ galaxies (that have 
{\it reference} but not necessarily {\it mean} luminosity for each 
morphological type), the natural expectation, when comparing with real 
galaxy samples, as in Fig. 5 and 6 for instance, is that models provide 
a sort of brighter (and somewhat redder) envelope to the data. 
This is because the real ``baricenter'' of the luminosity distribution 
at any $z$ is always slightly fainter (and bluer) than $L^*$, depending 
on the faint-end tail of LFs.

Even if the model is consistent with the observed color of galaxies
at bright magnitudes, it completely fails to match the observed trend 
at faint magnitudes where galaxies are much bluer than expected.  
The hypothesis of a high redshift galaxy population ($z>1$)
dominating the faint counts is thus unacceptable.
Even in the case that  Irregular galaxies would prevail the colors
should be  much redder than J-K$\sim1$.
Apparently, the equivalence {\em fainter magnitudes $\simeq$ 
higher redshifts} is therefore not supported by the observations.

\begin{figure}
\centerline{\psfig{figure=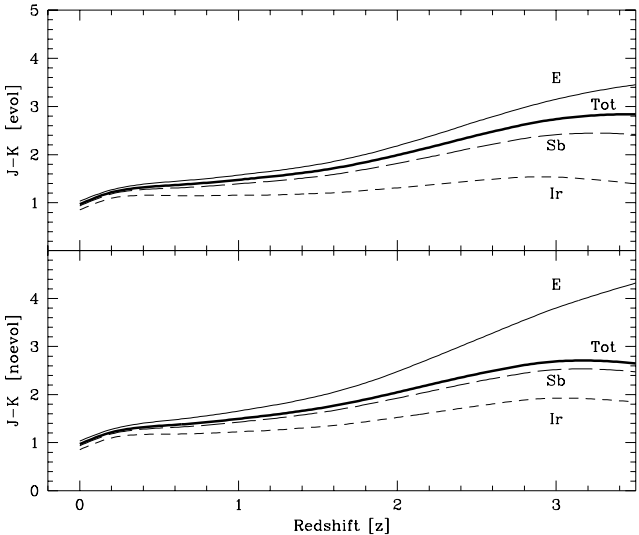,height=80mm,bbllx=40mm,bblly=80mm,bburx=180mm,bbury=250mm}}
\caption{Expected J-Ks color of different galaxy types as a function
 of redshift in the case of no-evolution (lower panel) and 
 evolution (upper panel).  The mean expected color of the local galaxy
 mix as in Marzke et al. (1998) is also shown (thick solid line).}
\end{figure}
A similar relationship between color trend and redshift distribution of faint
galaxies has also been found by Cowie et al. (1996). In their sample, 
they find  that blue objects dominate the K counts and that a huge fraction
of their K-selected faint objects are in fact extremely blue in (I-K).
Moreover, the observed median redshift of faint galaxies appears to 
be lower than 
expected for $L^*$ galaxies seen at the same apparent K magnitude, while 
galaxies display a roughly constant or even decreasing K  luminosity 
with increasing redshift at least at K$\le19$ where the redshift
completeness is high.

As  the high-reshift solution ($z>1$) seems to be inconsistent 
with the photometric properties of faint IR galaxies, we have to 
investigate the possibility of a local ($z<0.3$) origin of these galaxies.

A galaxy seen at Ks$\sim$22 and placed at $z\sim0.3$ would display an absolute
magnitude not brighter than $M_K\sim-19$ (assuming H$_0$=50 
Km  s$^{-1}$ Mpc$^{-1}$, q$_0$=0.5).
Integrating the  local IR luminosity function (LF) of 
galaxies in  Gardner et al. (1997) described by $\alpha\sim-1$ and
$M^*_K\sim-24.5$,
within  $z\sim0.3$ and  $M_K\ge-19$ we derive, in the magnitude
range  $21<K<22$, a surface density of galaxies
 20-40 times lower than observed.
A large uncertainty in such estimates derives from the different
values of the normalization parameter $\phi^*$ of the LF obtained by various
authors which can vary by more than a factor of two 
(see e.g. Efsthathiou et al. 1988; Lilly et al. 1995; 
Zucca et al. 1998, Mobasher et al. 1993; Glazebrook et al. 1995;
Gardner et al. 1997; Cowie et al. 1996).
Using a high normalization (e.g. $\phi^*\simeq0.005$ Mpc$^{-3}$ as in Marzke 
et al. 1994) the observed surface density of galaxies
can be obtained  within $z\sim0.3$ by assuming a slope of the faint end LF
$\alpha\le-1.8$. 

Some indications of a flat ($\alpha\sim-1$) faint-end LF in the near IR come
from the work of Gardner et al. (1997), Mobasher et al. (1993) and 
Glazebrook et al. (1995).
However, it is worth noting that all of the previous work has sampled 
the LF at relatively bright IR luminosity ($M_K<-20$) and thus misses the
crucial piece of information to assess our problem. 
On the contrary, some hints in favor of a steeper 
LF for field galaxies at infrared wavelengths come from  Szokoly et al. (1998) 
who find a slope $\alpha\sim-1.3$.
The small sample of galaxies used in that work (one fourth of that 
used by Gardner et al.) cannot, however, place severe constraints on the 
value of $\alpha$.
Stronger evidences of a steep K-band LF of field galaxies come from 
Bershady et al. (1999) which obtain $\alpha\sim-1.6$.

Evidence for a steeper slope of the IR LF are also found for cluster galaxies 
($\alpha\sim-1.4$, Mobasher and Trentham 1998; $\alpha\sim-2$,  
De Propris et al. 1998a) and, 
in the optical bands,  both in the cluster environment (e.g. De Propris et 
al. 1998b; Lobo et al. 1997; Bernstein et al. 1995; Driver et al. 1994;
Molinari et al. 1998)
and in the field{\footnote{In a recent paper Im et al. (1999) present a
morphologically divided redshift distribution of a sample of galaxies
limited to I=21.5. They argue that the faint end slope of the LF 
of very late type galaxies should be steeper than $\alpha=-1.4$}}: 
Marzke et al. (1994)  find $\alpha\sim-1.8$ for the 
LF of very late type and irregulars  galaxies (see also  Marzke and  
Da Costa 1997; Marzke et al. 1998); Zucca et al. (1998)  find a rising LF 
at magnitudes fainter than $M_{b_j}\sim-16$
which is well fitted by a power law with slope $\beta=-1.6$
and, finally, Liu et al. (1998) find $\alpha\sim-1.85$ for the U-band
LF of galaxies bluer than Sbc. 

Faced with the lack of reliable information on  deep IR LFs  in the 
field but with emerging evidence from 
optical investigations, we are inclined to believe  that 
a genuine steepening of the 
LF at its faint-end tail cannot be ruled out and that 
a prevailing population of sub-L$^*_K$ (and hence low mass and low
redshift ($z<1$)) 
galaxies may dominate at faint infrared magnitudes.

\subsection{Sizes, Compactness and Galaxy Counts}
From Figs. 7 and 8  we can infer that the faint IR population mainly consists
of galaxies which are intrinsically compact compared to those seen 
at brighter apparent magnitudes.
Fig. 9 also shows that the contribution of small,  compact galaxies 
to the counts increases faster than that of large galaxies going to 
faint apparent magnitudes as shown by the steeper slopes derived.
It is worth noting that the counts of small sources follow an Euclidean slope
which indicates that either they are very low redshift galaxies, very rapidly
evolving or a combination of the two.
Finally, Figs. 10, 11 and 12 indicate that small galaxies 
contribute most to the blue population.
{\it Our  conclusion is that 
sub-L$^*_K$ galaxies contribute to the number counts and with increasing 
weight to fainter magnitudes.}

At  faint apparent magnitudes we are sampling the faint-end tail
of the local luminosity function superposed on the bright tail ($M_K\le M^*_K$)
of the high redshift galaxy population.
The relative contribution to the total number of galaxies at a given
apparent magnitude depends on the volume sampled by a given luminosity
and on the density of galaxies with that luminosity.
A large contribution to faint counts could come from the faint end of the LF
in the case of a steep slope ($\alpha<-1$) since the number of luminous 
galaxies would increase slower than that of the intrinsically faint ones
at faint apparent magnitudes.
In this case we should first observe  a reddening trend of the J-K color
due to the rapid increase of galaxies with  progressively higher redshift
going to fainter apparent magnitudes.
Such a trend would be expected to continue steadily as long as the
number of intrinsically faint galaxies, and hence low redshift
galaxies, starts to increase more rapidly than high-$z$ galaxies.  
The mean observed color should therefore become increasingly bluer,
contrary to what is expected for  the standard case of a normal
high-$z$ galaxy population.  Moreover, the median redshift would
increase more slowly than expected going to fainter apparent
magnitudes, as observed by Cowie et al. (1996) in their redshift
survey.

For example, assuming a slope $\alpha=-1.6$ at luminosities
$L<0.01L^*$, as in the ESP (Zucca et al. 1997), by integrating the LF
down to 0.0001$L^*$ we find that more than 40$\%$ of the galaxies in
the magnitude range $21<K<22$ are at $z<0.5$ and more than 60$\%$ at
$z<1$.  These fractions could even increase to 60$\%$ and 80$\%$ in
the case of a steeper LF, for example according to the result of
Marzke et al. (1994, 1998) or De Propris et al. (1998), or integrating
the LF down to a fainter cutoff.  It is worth noting that this simple
scenario, which is a straightforward consequence of a steepening trend
in the LF faint-end tail, would also account for the observed trend in
galaxy size and compactness parameter, as discussed earlier.  Assuming
that faint galaxies are actually intrinsically faint IR objects and
that the IR luminosity is also a fair tracer of mass (Gavazzi et
al. 1996) implies that compact systems of low mass constitute an
important population in the Universe at least back to $z\sim 1$.

\section{Summary, Conclusions and Future Work}
We have presented counts, colors and sizes of galaxies detected in
deep J (down to J=24) and Ks (down to Ks=22.5) images obtained with
the new IR imager/spectrometer SOFI at the ESO NTT telescope.  These
data represent the largest samples collected so far reaching these
depths.  The main results obtained from analysis of the data are the
following:
\begin{itemize}
\item number counts follow a  $d~log(N)/dm$ relation with slope 
 0.38 in Ks and  0.36 in J showing no sign of any flattening or 
turnover down to the faintest magnitudes. 
This fully agrees with most of the previous ground-based 
data and with the deepest NICMOS/HST data; 

\item fainter than Ks$\sim19$ and J$\sim20$, the median J-K galaxy
color  shows a break in its reddening trend and turns toward bluer colors;

\item faint bluer galaxies display both a larger compactness index  
and smaller apparent sizes;

\item the J-K color distribution of small and of large galaxies 
(selected with respect to the median size of the whole sample) 
are significantly 
different with small galaxies contributing most to the 
blue population;

\item the number counts of small galaxies follow an Euclidean slope,
 much steeper than that of large sources, and dominate the  deep number 
counts fainter than Ks=21.5 and J=23.

\end{itemize}
The absence of a turnover in the counts down to the faintest magnitudes
is indicative of an increasing contribution of sub-L$^*$  
and thus low redshift galaxies ($z<1$) 
leading to envisage a substantial steepening in the  faint-end tail 
of the galaxy LF. 
Such a claim is also supported by the observed color trend which 
seems to rule out
any major contribution of high-$z$ galaxies as primary contributors 
to faint counts. 
Moreover also the Euclidean slope shown by counts of small galaxies
supports this claim.
In particular the Euclidean slope indicates that either small galaxies
 are at very low redshift, or that they are very 
rapidly evolving or a combination of the two.

The existence of a steep  ($\alpha\ll-1$) faint-end tail ($L<0.1-0.01L^*$)
in the IR LF would naturally account both for the J-K color trend,
the compactness trend and the different slope shown by the counts
of small and large galaxies, the former being described by a much steeper
(nearly Euclidean) slope.

Very little is known about the redshift distribution of galaxies to
faint K magnitudes.  Two spectroscopic surveys, Cowie et al. (1996)
and Cohen et al. (1999) based on a K=20 limited sample have been
carried out so far.  Taking into account their redshift
incompleteness, it seems that 10-30$\%$ of objects are at $z>1$.  The
population we are discussing here is concentrated at K$>20$.  It will
prove quite difficult to obtain spectroscopic redshifts for these
objects since their $R$ magnitudes should be fainter than 24 and
obtaining IR spectra would be equally difficult even using an IR
spectrometer like ISAAC at the VLT.  The most promising approach to
understand the properties of the faint population is the use of
photometric redshifts based on color measurements from the U to K
bands (Fontana et al. 1999a; Benitez et al. 1999).  This is already
being done on the SUSI NTT Deep Field, a subsection of the area
covered by the images analyzed in this paper (Arnouts et al. 1999;
Fontana et al. 1999b).

\begin{acknowledgements}
We wish to thank the referee, M. Bershady, for the valuable comments
and suggestions which significantly improved the manuscript.  
The authors are grateful to N. Devillard for his valuable help in
the data reduction and P. A. Duc for providing support in the astrometric 
calibration.
PS is grateful to A. Iovino and G. Chincarini for helpful discussions
and their suggestions.

\end{acknowledgements}

\end{document}